\documentclass[onecolumn,preprintnumbers,aps,superscriptaddress,nofootinbib,prd,notitlepage]{revtex4-1}
\pdfoutput=1 

\usepackage{color}
\usepackage{amsmath}
\usepackage{graphicx}
\usepackage{scalefnt}

\usepackage{grffile} % allow for more than 1 '.' in file name for figure

\newcommand{\Lmu}{L_\mu}

\graphicspath{ {figs/} }

\allowdisplaybreaks

\begin{document}

\preprint{P3H-22-026, TTP22-014}

\title{Three-loop non-singlet matching coefficients for heavy quark currents}
\author{Manuel Egner}
\email{manuel.egner@kit.edu}
\affiliation{Institut f\"ur Theoretische Teilchenphysik,
  Karlsruhe Institute of Technology (KIT), 76128 Karlsruhe, Germany}
\author{Matteo Fael}
\email{matteo.fael@kit.edu}
\affiliation{Institut f\"ur Theoretische Teilchenphysik,
  Karlsruhe Institute of Technology (KIT), 76128 Karlsruhe, Germany}
\author{Fabian Lange}
\email{fabian.lange@kit.edu}
\affiliation{Institut f\"ur Theoretische Teilchenphysik,
  Karlsruhe Institute of Technology (KIT), 76128 Karlsruhe, Germany}
\affiliation{Institut f{\"u}r Astroteilchenphysik,
  Karlsruhe Institute of Technology (KIT), 76344 Eggenstein-Leopoldshafen, Germany}
\author{Kay Sch\"onwald}
\email{kay.schoenwald@kit.edu}
\affiliation{Institut f\"ur Theoretische Teilchenphysik,
  Karlsruhe Institute of Technology (KIT), 76128 Karlsruhe, Germany}
\author{Matthias Steinhauser}
\email{matthias.steinhauser@kit.edu}
\affiliation{Institut f\"ur Theoretische Teilchenphysik,
  Karlsruhe Institute of Technology (KIT), 76128 Karlsruhe, Germany}
%\date{}

\begin{abstract}
  We compute the matching coefficients between QCD and
  non-relativistic QCD for external vector, axial-vector, scalar and
  pseudo-scalar currents up to three-loop order. We concentrate on the
  non-singlet contributions and present precise numerical results
  with an accuracy of about ten digits. For the vector current
  the results from Ref.~\cite{Marquard:2014pea} are confirmed,
  increasing the accuracy by several orders of magnitude.
\end{abstract}

%\pacs{}
\maketitle

%- }}}
%- {{{ Intro.:

\section{Introduction}

The construction of effective field theories with Quantum Chromodynamics
(QCD) as a starting point is a very successful approach in order to
describe a number of different phenomena, which involve different energy
scales following a large hierarchy.
A popular example in this context is non-relativistic QCD (NRQCD) which
describes systems with two heavy quarks moving with small relative
velocity. Prominent applications are the threshold production of top-quark
pairs in electron-positron annihilation and properties of charmonium and
bottomonioum bound states. For comprehensive reviews we refer to
Refs.~\cite{Pineda:2011dg,Beneke:2013jia}.

For the construction of the effective theories one considers Green functions
in the full and effective theories and requires that they are equal up to
corrections in the small expansion parameter, which in the case of NRQCD are
power-suppressed terms in the inverse heavy quark mass $m$.  Such
calculations, usually referred to as matching calculations, fix the couplings of
the operators in the effective theory. These couplings are typically denoted as
matching coefficients.

\begin{figure}[t]
  \begin{center}
    \begin{tabular}{cccc}
      \includegraphics[width=.2\textwidth]{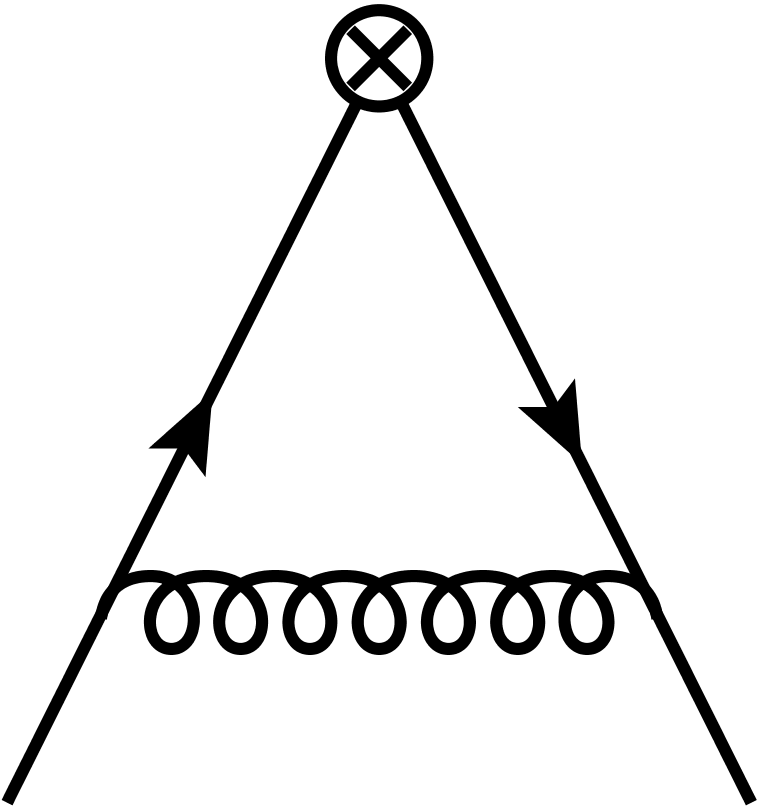} &
      \includegraphics[width=.2\textwidth]{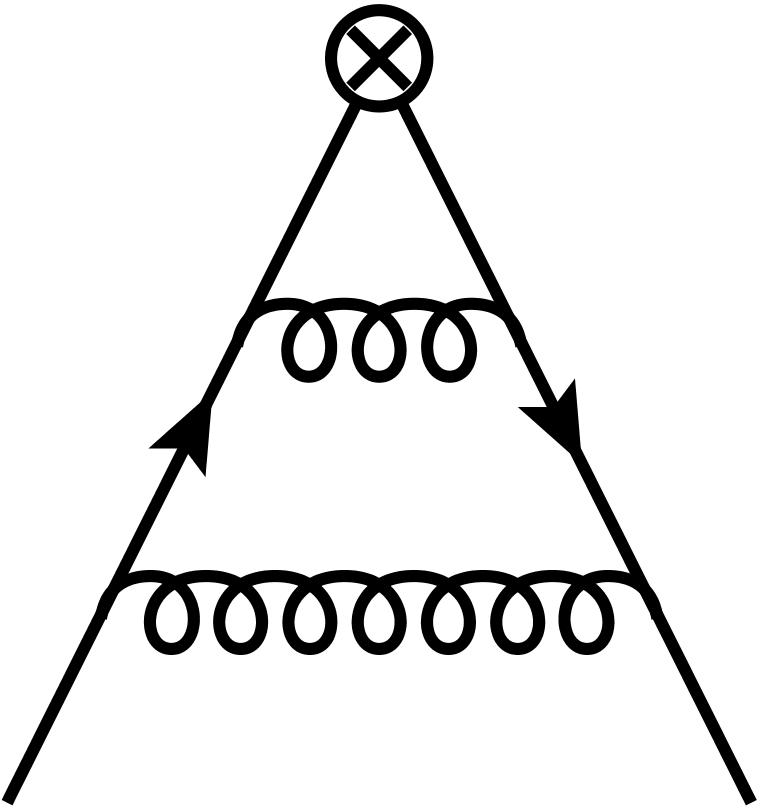} &
      \includegraphics[width=.2\textwidth]{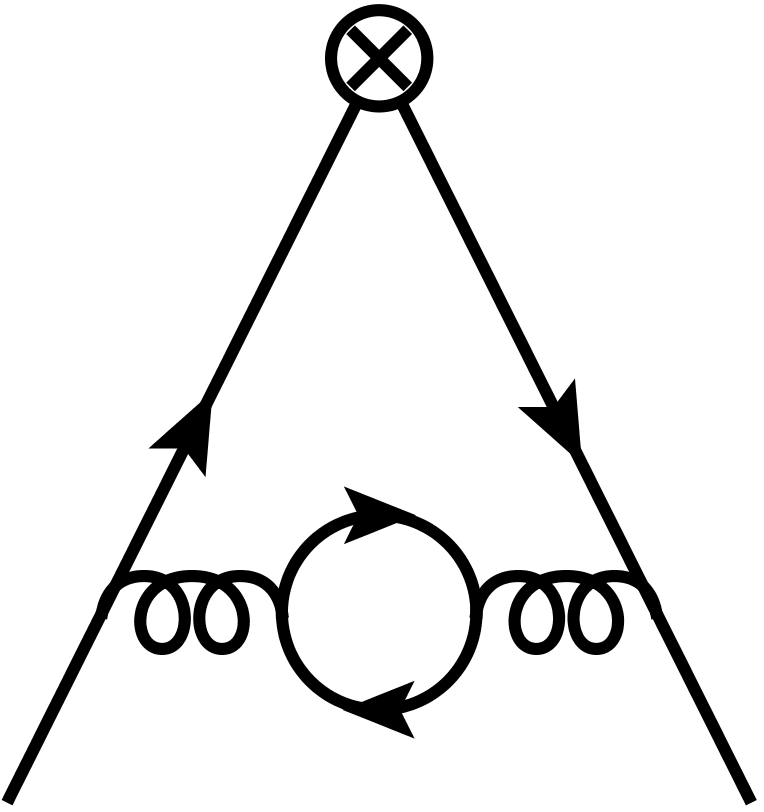} &
      \includegraphics[width=.2\textwidth]{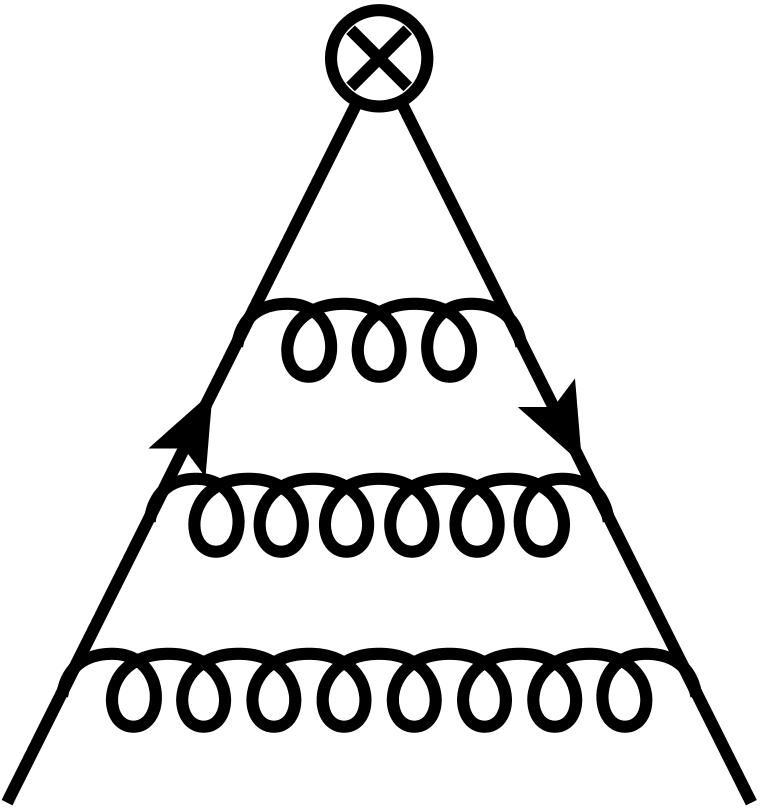} \\
      (a) & (b) & (c) & (d) \\
      \includegraphics[width=.2\textwidth]{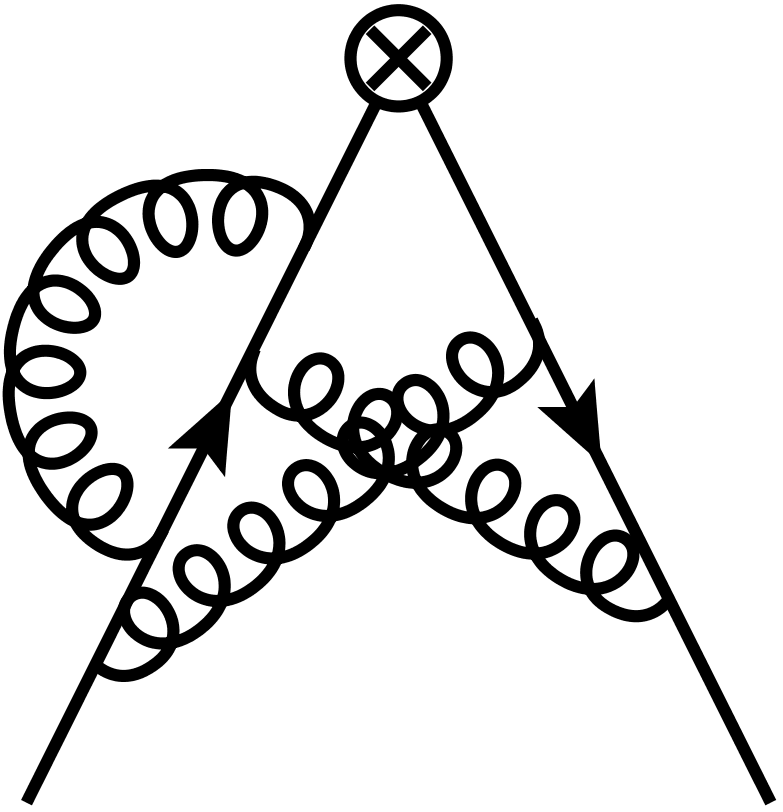} &
      \includegraphics[width=.2\textwidth]{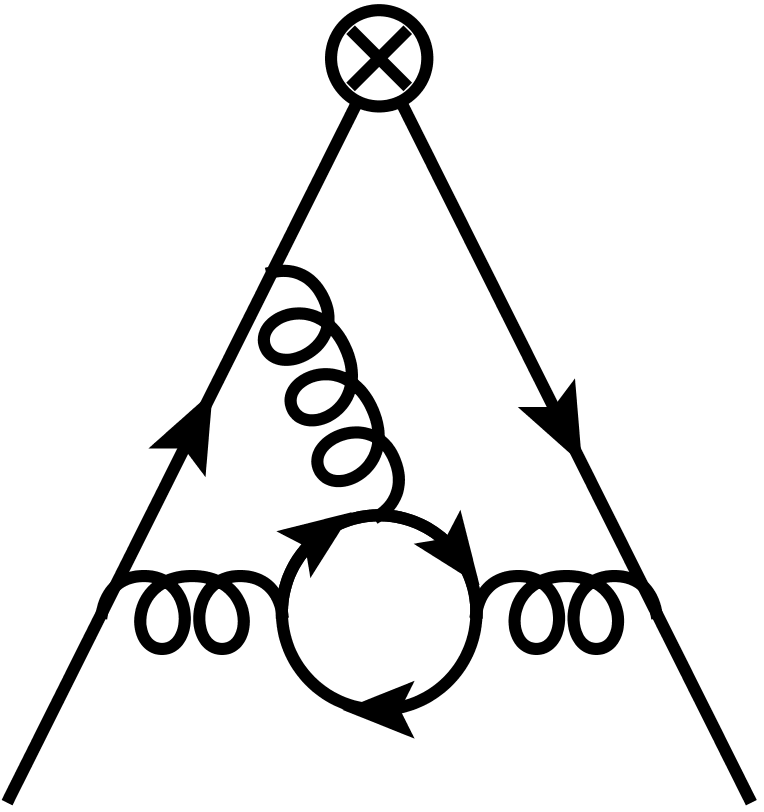} &
      \includegraphics[width=.2\textwidth]{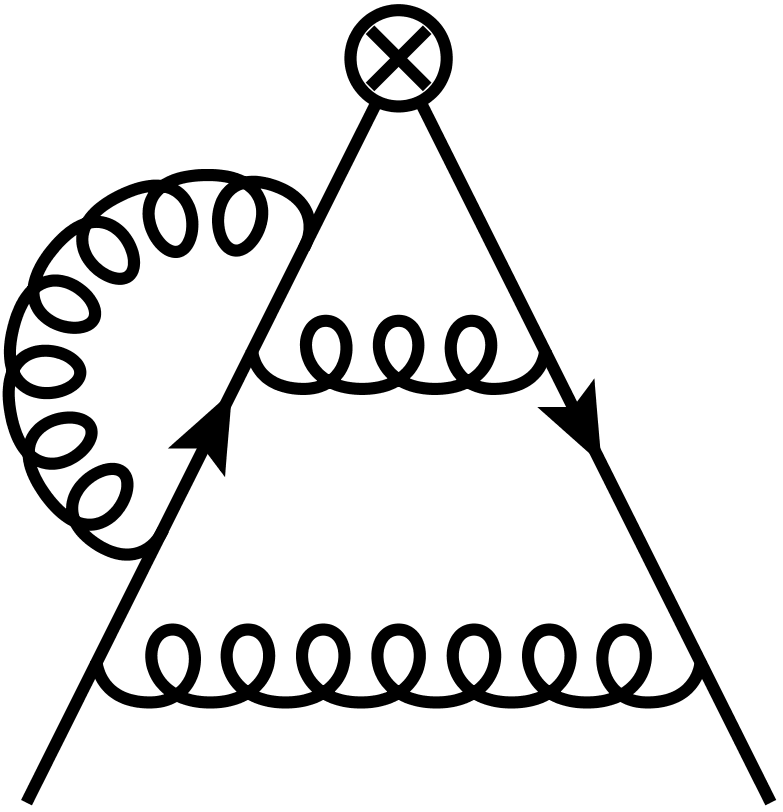} &
      \includegraphics[width=.2\textwidth]{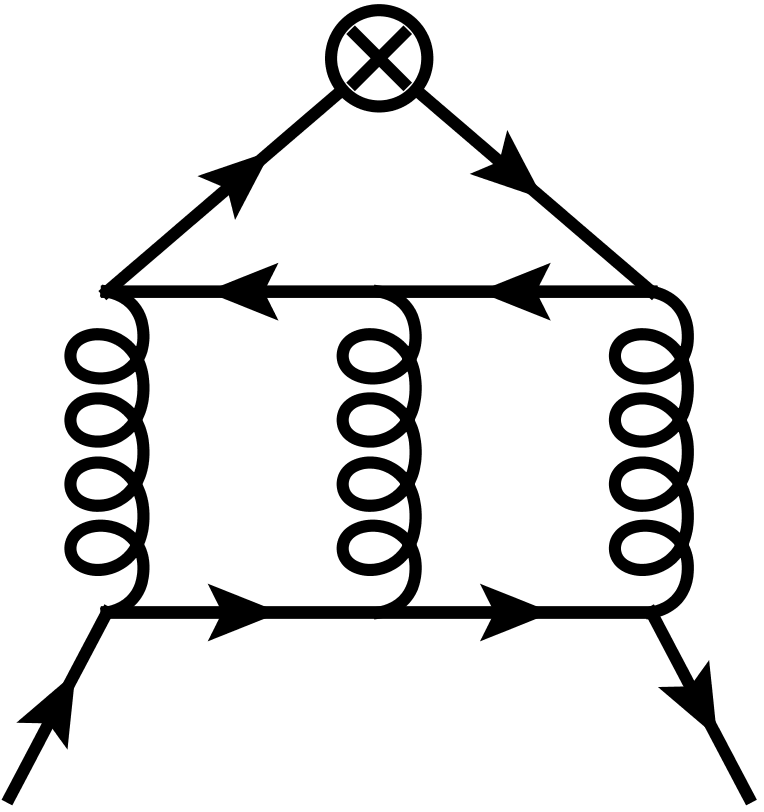} \\
      (e) & (f) & (g) & (h)
    \end{tabular}
    \caption{\label{fig::diags}Sample Feynman diagrams at one-, two- and
      three-loop order for the current-quark-anti-quark vertex corrections.
      Solid and curly lines denote quarks and gluons,
      respectively. The cross represents the coupling to the external current.
      In this work we only consider non-singlet contributions (a)-(g)
      and neglect the singlet contributions (h).}
  \end{center}
\end{figure}

In this paper we consider QCD and NRQCD as full and effective theories and
compute the matching coefficients of external vector, axial-vector, scalar and
pseudo-scalar currents up to three-loop order in perturbation theory.  For
this purpose it is necessary to compute vertex corrections involving one of
the currents and a quark-anti-quark pair.  We concentrate on the
non-singlet contributions where the external currents directly couple to the
external quarks. Sample Feynman diagrams up to three loops are shown in
Fig.~\ref{fig::diags}.

From the phenomenological point of view the vector current is certainly most
important. It enters as building block to the threshold production of top-quark
pairs~\cite{Beneke:2015kwa} and the decay width of the $\Upsilon(1S)$
meson~\cite{Beneke:2014qea,Egner:2021lxd}. Its abelian contribution
is an important ingredient to the hyperfine splitting of
positronium~\cite{Baker:2014sua}. As possible applications of the
scalar and pseudo-scalar matching coefficient one could imagine
the decay of CP-even or CP-odd Higgs bosons with mass $M$ into two quarks
with mass $m\approx M/2$.

Starting point for the matching calculation are the vector, axial-vector,
scalar and pseudo-scalar currents in QCD which we define as
\begin{eqnarray}
  j_v^\mu &=& \bar{\psi} \gamma^\mu \psi\,, \nonumber \\
  j_a^\mu &=& \bar{\psi} \gamma^\mu\gamma_5 \psi\,, \nonumber \\
  j_s     &=& \bar{\psi} \psi\,, \nonumber \\
  j_p     &=& \bar{\psi} {\rm i}\gamma_5 \psi\,.
  \label{eq::currents}
\end{eqnarray}
Note that the anomalous dimensions of the vector and axial-vector current
are zero whereas $j_s$ and $j_p$ involve non-trivial renormalization constants.

Expanding the spinors in Eq.~(\ref{eq::currents}) for $|\vec{p}\,|\ll m$, 
where $\vec{p}$ is the momentum of the anti-quark in the final state, one
finds the currents in the effective theory,
\begin{eqnarray}
  \tilde{j}_v^k &=& \phi^\dagger \sigma^k \chi \,,\nonumber\\
  \tilde{j}_a^k &=& \frac{1}{2 m} \phi^\dagger
  [\sigma^k,\vec{p}\cdot\vec{\sigma}] \chi \,,\nonumber\\
  \tilde{j}_s &=& -\frac{1}{m} \phi^\dagger \vec{p}\cdot\vec{\sigma}
  \chi \,,\nonumber\\
  \tilde{j}_p &=& -{\rm i}\phi^\dagger \chi \,,
                  \label{eq::currents_eff}
\end{eqnarray}
where $\phi$ and $\chi$ are two-component Pauli spinors.

The currents in Eqs.~(\ref{eq::currents}) and~(\ref{eq::currents_eff}) are
used to form renormalized vertex functions with two external on-shell quarks
which we denote by $\Gamma_x(q_1,q_2)$ and $\tilde{\Gamma}_x$ with
$x\in\{v,a,s,p\}$, respectively.  $q_1$ and $q_2$ correspond to
the momenta of the quark and anti-quark with $q_1^2=q_2^2=m^2$ where $m$ is
the quark mass.  We apply an asymptotic expansion around
$s=4m^2$~\cite{Beneke:1997zp,Smirnov:2002pj}, where $s$ is the momentum
squared of the external current, which leads to
\begin{eqnarray}
  Z_2 Z_x \Gamma_x(q_1,q_2) &=& c_x \tilde{Z}_2 \tilde{Z}_x^{-1}
  \tilde{\Gamma}_x + \ldots
  \label{eq::match_def}
  \,.
\end{eqnarray}
The ellipses denote terms suppressed by at least two inverse powers of
the heavy quark mass. 
It is understood that $\Gamma_x(q_1,q_2)$ is expressed in terms of the
heavy quark mass in the on-shell scheme and the strong coupling in the
$\overline{\rm MS}$ scheme.
$Z_2$ and $\tilde{Z}_2$ are the on-shell wave function
renormalization constants. $Z_2$ is needed up to three
loops~\cite{Melnikov:2000zc,Marquard:2007uj} whereas $\tilde{Z}_2=1$ since the
quantum corrections in NRQCD only involve scaleless integrals which are set to
zero in dimensional regularization. Also for $\tilde{\Gamma}_x$ only
tree-level contributions are needed since the soft, potenial and ultrasoft contributions
are present on both sides of Eq.~(\ref{eq::match_def}) 
and cancel such that only the hard contribution of $\Gamma_x(q_1,q_2)$ has to
be computed. $Z_x$ is the renormalization constant of the current in
full QCD which is given by $Z_v=Z_a=1$ and $Z_s=Z_p=Z_m$. Here $Z_m$ is the
on-shell quark mass renormalization constant defined via $m = Z_m m^0$, where
$m_0$ is the bare heavy quark mass.  $\tilde{Z}_x$ is the renormalization
constant of the current in NRQCD which is determined from the infrared
divergences of $c_x$.  $\tilde{Z}_x$ deviates from 1 starting at order
$\alpha_s^2$.  The computation of the matching coefficient $c_x$ is the main
purpose of this work.  

Two-loop corrections to $c_v$ have been computed for the first time in
Refs.~\cite{Czarnecki:1997vz,Beneke:1997jm} and in Ref.~\cite{Kniehl:2006qw}
two-loop corrections to all four currents have been considered, including the
singlet contributions. Three-loop corrections to $c_v$ have been computed in
Refs.~\cite{Marquard:2006qi,Marquard:2009bj,Marquard:2014pea}. In these works
the reduction to master integrals has been performed analytically.
However, most of the master integrals have only been computed numerically with the help of
{\tt FIESTA}~\cite{Smirnov:2015mct}. As a consequence the coefficients of some colour
structures are only know with an uncertainty of a few percent.
This is sufficient for most phenomenological applications. It is nevertheless
desirable to have an independent cross check with
improved accuracy. This is provided in this work.

In the next Section
we provide details on our calculation and describe our method to extract
the matching coefficient from results for the form factors.  In
Section~\ref{sec::results} we present our results for the matching coefficients
and the anomalous dimension of the currents in the effective theory.
Section~\ref{sec::concl} contains a brief summary.

%- }}}
%- {{{ Technical details:

\section{Technical details}

For the computation of the hard part of the vertex diagrams we apply the
method developed in Ref.~\cite{Fael:2021kyg}. We profit from the findings of
Refs.~\cite{Fael:2022rgm,FLSS22} where results for massive form factors with
external vector, axial-vector, scalar and pseudo-scalar currents have been
computed. They can be decomposed into six form factors given by
\begin{eqnarray}
  \Gamma_\mu^v(q_1,q_2) &=&
  F_1^v(s)\gamma_\mu - \frac{\mathrm{i}}{2m}F_2^v(s) \sigma_{\mu\nu}q^\nu
  \,, \nonumber\\
  \Gamma_\mu^a(q_1,q_2) &=&
  F_1^a(s)\gamma_\mu\gamma_5 {- \frac{1}{2m}F_2^a(s) q_\mu }\gamma_5
  \,, \nonumber\\
  \Gamma^s(q_1,q_2) &=& {m} F^s(s)
  \,, \nonumber\\
  \Gamma^p(q_1,q_2) &=& {\mathrm{i} m} F^p(s) {\gamma_5}
  \,,
  \label{eq::Gamma}
\end{eqnarray}
where $\sigma^{\mu\nu} = \mathrm{i}[\gamma^\mu,\gamma^\nu]/2$ and
$s$ is the invariant mass of the external current. The quantity
$\Gamma_x(q_1,q_2)$ in Eq.~(\ref{eq::match_def}) is obtained from the hard
part of the form factors evaluated at $s=4m^2$ through
\begin{eqnarray}
  \Gamma_v &=& (F_1^v + F_2^v)\big|_{\rm hard,\, s=4m^2}\,,\nonumber\\
  \Gamma_a &=& F_1^a\big|_{\rm hard,\, s=4m^2}\,,\nonumber\\
  \Gamma_s &=& F^s\big|_{\rm hard,\, s=4m^2}\,,\nonumber\\
  \Gamma_p &=& F^p\big|_{\rm hard,\, s=4m^2}\,,
            \label{eq::c_from_F}
\end{eqnarray}
which is discussed in more detail in the remainder of this section.

The basic idea of Ref.~\cite{Fael:2021kyg} is to construct expansions of the master
integrals for various values of $s/m^2$ with the help of the corresponding
differential equations. The unconstrained coefficients of the expansions are
fixed by matching two neighboring expressions at an intermediate point.  The
starting point in Refs.~\cite{Fael:2022rgm,FLSS22} is $s=0$ where all master
integrals can be computed analytically. In order to arrive at the threshold
$s=4m^2$ we perform expansions for $s/m^2 = 1,2,5/2,3,7/2$ and
$4$.

The expansion around $s/m^2=4$ uses the variable
\begin{align}
        x &= \sqrt{ 4 - \frac{s}{m^2} }\,.
\end{align}
It contains both even and odd powers of $x$ accompanied by $\ln(x)$ terms,
since it comprises the contributions from all regions present close to
threshold.  In particular, each loop momentum can have one of the following
scalings~\cite{Beneke:1997zp}:\footnote{Note that in Ref.~\cite{Beneke:1997zp}
  the variable $y=1-s/(4m^2) = x^2/4$ has been used.}
\begin{itemize}
        \item hard (h): $k_0 \sim m$, $k_i \sim m$ ,
        \item potential (p): $k_0 \sim x^2 \cdot m$, $k_i \sim x \cdot m$ ,
        \item soft (s): $k_0 \sim x \cdot m$, $k_i \sim x \cdot m$ ,
        \item ultrasoft (u): $k_0 \sim x^2 \cdot m$, $k_i \sim x^2 \cdot m$ .
\end{itemize}
For the matching coefficients we only need the region where
all loop momenta are hard. Here only even powers of $x$ and
no $\ln(x)$ terms are present.

Using the scalings from above, we see that in each region the integral is
given as $x^{-n \, \epsilon}$ multiplied by a Taylor expansion in $x$, with an
integer $n$ which can be derived from the scaling of the loop momenta in the
respective region.  Here $\epsilon=(4-d)/2$ where $d$ is the space-time
dimension.  We can insert this ansatz into the system of differential
equations for the master integrals and obtain a system of linear equations for
the expansion coefficients.  For each region the system is reduced to a small
set of undetermined boundary constants with the help of a version of
\texttt{Kira}~\cite{Maierhofer:2017gsa,Klappert:2020nbg} with
\texttt{FireFly}~\cite{Klappert:2019emp,Klappert:2020aqs} optimized for
solving systems without variables.  After summing the contributions from all
regions we obtain again the results for the master integrals in full
kinematics.  We can therefore numerically match the yet undetermined boundary
constants with the numerical results computed in Ref.~\cite{Fael:2022rgm}.
Substituting the numerical solutions into the ansatz for the
$x^{-0 \, \epsilon}$ scaling provides the master integrals in the hard
expansion.

Let us in the following discuss the calculation in more detail.
At two-loop order we find the following scalings for the different regions:
\begin{itemize}
        \item $x^{-0 \epsilon}$: (h-h),
        \item $x^{-2 \epsilon}$: (h-p), (h-s),
        \item $x^{-4 \epsilon}$: (h-u), (p-p), (s-s), (p-s),
        \item $x^{-6 \epsilon}$: (p-u), (s-u),
        \item $x^{-8 \epsilon}$: (u-u),
\end{itemize}
where the list on the right of the colon specifies the scaling of the two loop
momenta. Some of the combinations might vanish due to the presence
of scaleless integrals. However, in our approach we do not have to pay
attention to this.  Since only the spacial parts get continued into
$(d-1)$ dimensions, potential and soft regions of the loop momenta lead to
the same $\epsilon$-dimensional scalings.  The pure ultrasoft region
$\sim x^{-8 \epsilon}$ does not contribute which we checked by an explicit
calculation.  For the two-loop calculation we therefore have to consider four
independent expansions.  Note that the individual regions contributing to one
of the $x^{-n \epsilon}$ scalings might develop higher poles in the
dimensional regulator $\epsilon$ than the sum.  These higher poles lead to
Sudakov-like double logarithms which are not present in the threshold
expansion considered here.  We therefore do not have to extend the ansatz to
higher poles in $\epsilon$ compared to the full calculation in
Ref.~\cite{Fael:2022rgm}.

At three loop order we have the scalings
\begin{itemize}
        \item $x^{-0 \epsilon}$: (h-h-h),
        \item $x^{-2 \epsilon}$: (h-h-p), (h-h-s),
        \item $x^{-4 \epsilon}$: (h-h-u), (h-p-p), (h-s-s), (h-p-s),
        \item $x^{-6 \epsilon}$: (h-p-u), (h-s-u), (p-p-p), (p-p-s), (p-s-s), (s-s-s),
        \item $x^{-8 \epsilon}$: (h-u-u), (u-p-p), (u-p-s), (u-s-s),
        \item $x^{-10 \epsilon}$: (u-u-p), (u-u-s),
        \item $x^{-12 \epsilon}$: (u-u-u),
\end{itemize}
which means that we have to construct six independent expansions since the
pure-ultrasoft contribution vanishes.  After the reduction to boundary
constants we are left with $(568,125,248,402,236,51)$ undetermined
coefficients for the scalings $x^{-0 \epsilon}, \ldots, x^{-10 \epsilon}$.
They can be reduced by utilizing information about the master integrals from
the full calculation.  On the one hand, we know some integrals analytically,
especially those which do not depend on $s$. They can be fixed from the
expansion around $s=0$. Furthermore, some of the $\epsilon$ poles also do not
have a $s$ dependence and thus also they are available from the calculation
performed for $s=0$.  On the other hand, we know the leading power in $x$ for
each integral from the full result.  This knowledge implies relations between
the boundary constants from different regions which leads to a reduction of
the number of independent boundary constants from 1630 to 578.  They are
determined as follows: After obtaining the symbolic expansions for each region
we equate the sum of all regions with the numerical evaluation of the full
result at $s=3.75 m^2$ from Ref.~\cite{Fael:2022rgm} and solve the resulting linear
system for the 578 boundary constants.  In particular all 568 coefficients
from the pure-hard regions of all 422 master integrals are obtained by this
procedure, whereas the regions which scale as $x^{-n \epsilon}$ with $n>0$ can
not be disentangled. This is sufficient for the application in the present
paper.

Let us mention that in case one wants to construct results for each individual
region further information is needed. It can be obtained by determining for
each region of every master integral the leading power in $x$.  Here the
program \texttt{asy.m}~\cite{Pak:2010pt,Jantzen:2012mw} can be used.  In this
way one obtains relations for each individual region instead of only for the
sum of all of them.

Next we insert the hard regions of the master integrals into the amplitudes
for the form factors.  It contains terms scaling with inverse powers of
$(s-4m^2)$ from the reduction of the master integrals with full
kinematics.  It is a non-trivial check that the limit $s\to 4m^2$ exists.  In
fact we have checked that all inverse powers of $(s-4m^2)$ have coefficients
below $3 \cdot 10^{-11}$  which is the precision of our calculation.
Inserting the form factors into Eq.~(\ref{eq::c_from_F}) we finally obtain the vertex
functions $\Gamma_x$ entering the matching equation~(\ref{eq::match_def}).
As a further check we keep the QCD gauge parameter $\xi$ and observe that it
vanishes after renormalization.

%- }}}
%- {{{ Results:

\section{\label{sec::results}Three-loop matching coefficients}

Once all ingredients for the left-hand-side of Eq.~(\ref{eq::match_def}) are
available we can solve it for $c_x$ order-by-order in $\alpha_s$.  At one-loop
order all quantities with a tilde on the right-hand-side are equal to 1. At
order $\alpha_s^2$ infrared divergences are left on the left-hand-side which are
absorbed into $\tilde{Z}_x$. Finally, at order $\alpha_s^3$ one has to take care of
the interference term of $\tilde{Z}_x^{-1}$ and the one-loop result of $c_x$,
which is needed up to order $\epsilon$. The remaining infrared divergences are again
absorbed into $\tilde{Z}_x$.
We parametrize the perturbative results in this section by the strong coupling
in the effective theory with $n_l$ active quark flavours which we denote by $\alpha_s^{(n_l)}$.

Let us in a first step provide the results for the renormalization constants
which are obtained by subtracting the remaining infrared divergences
in a minimal way. For the vector current we have
\begin{eqnarray}
  \tilde{Z}_v &=&
  1 + \left(\frac{\alpha_s^{(n_l)}(\mu)}{\pi}\right)^2
  \frac{C_F\pi^2}{\epsilon}\left(
    \frac{1}{12} C_F + \frac{1}{8} C_A \right)
%  \nonumber \\ &&\mbox{}
  + \left(\frac{\alpha_s^{(n_l)}(\mu)}{\pi}\right)^3  C_F\pi^2
   \Bigg\{ C_F^2 \left[\frac{5}{144\epsilon^2} + \left(
    \frac{43}{144} - \frac{1}{2} l_2 + \frac{5}{48} \Lmu
    \right)\frac{1}{\epsilon} \right] \nonumber \\ &&\mbox{}
  +C_FC_A \left[\frac{1}{864\epsilon^2} + \left( \frac{113}{324} +
    \frac{1}{4} l_2 + \frac{5}{32} \Lmu \right) \frac{1}{\epsilon}
    \right] %\nonumber \\ &&\mbox{}
  +C_A^2 \left[-\frac{1}{16\epsilon^2} + \left( \frac{2}{27} +
    \frac{1}{4} l_2 + \frac{1}{24} \Lmu \right) \frac{1}{\epsilon}
    \right] \nonumber \\ &&\mbox{}
  + T n_l \left[ C_F\left(
      \frac{1}{54\epsilon^2}
      -\frac{25}{324\epsilon}
    \right)
    + C_A \left(
      \frac{1}{36\epsilon^2}
      - \frac{37}{432\epsilon} \right)
     \right] %\nonumber \\ &&\mbox{}
  + C_F T n_h \frac{1}{60\epsilon}
  \Bigg\} + {\cal O}(\alpha_s^4)\,,
  \label{eq::Zv}
\end{eqnarray}
which agrees with the explicit calculations in the effective theory from
Refs.~\cite{Beneke:1997jm,Marquard:2006qi,Kniehl:2002yv,Beneke:2007pj}.
In Eq.~(\ref{eq::Zv}) $C_F=(N_c^2-1)/(2N_c)$ and $C_A=2TN_c$ are the quadratic Casimir operators of
the $\mathrm{SU}(N_c)$ gauge group in the fundamental and adjoint representation,
respectively,  $n_l$ is the number of massless quark flavors, and $T=1/2$.
Furthermore we have $\Lmu=\ln(\mu^2/m^2)$ and $l_2=\ln(2)$.

For the remaining three currents our results read
\begin{eqnarray}
  \tilde{Z}_a &=&
  1
  + \left(\frac{\alpha_s^{(n_l)}(\mu)}{\pi}\right)^2 \frac{C_F \pi^2}{\epsilon}
  \left(
    \frac{1}{24} C_A
    +\frac{5}{48} C_F
  \right)
  + \left(\frac{\alpha_s^{(n_l)}(\mu)}{\pi}\right)^3 C_F \pi^2
  \Biggr\{
    C_F^2 \left(\frac{215}{864}-\frac{l_2}{3}\right) \frac{1}{\epsilon }
    \nonumber \\ &&
    + C_F  C_A
    \left[
      -\frac{25}{576 \epsilon ^2}
      + \left(
      \frac{1}{18} l_2
      +\frac{35}{576} L_{\mu }
      +\frac{1433}{5184}
      \right) \frac{1}{\epsilon }
    \right]
    +C_A^2
    \left[
      -\frac{1}{48 \epsilon ^2}
      + \left(
        \frac{5}{36} l_2
        +\frac{1}{72} L_{\mu }
        +\frac{17}{324}
      \right) \frac{1}{\epsilon }
    \right]
    \nonumber \\ &&
    + T n_l
    \left[
      C_F
      \left(
        \frac{5}{216 \epsilon ^2}-\frac{83}{1296 \epsilon }
      \right)
      + C_A
      \left(
        \frac{1}{108 \epsilon ^2}-\frac{53}{1296 \epsilon }
      \right)
    \right]
  \Biggr\}
  \,,\nonumber\\
  \tilde{Z}_s &=&
  1
  + \left(\frac{\alpha_s^{(n_l)}(\mu)}{\pi}\right)^2 \frac{C_F \pi^2}{\epsilon}
  \left(
    \frac{1}{24}  C_A
    +\frac{1}{6}  C_F
  \right)
  + \left(\frac{\alpha_s^{(n_l)}(\mu)}{\pi}\right)^3 C_F \pi^2
  \Biggr\{
    C_F^2
    \left(
      \frac{65}{216}-\frac{1}{3}l_2
    \right)\frac{1}{\epsilon }
    \nonumber \\ &&
    + C_F C_A
    \left[
      -\frac{7}{96 \epsilon ^2}
      + \left(
         \frac{461}{1296}
        +\frac{1}{18} l_2
        +\frac{25}{288} L_{\mu }
      \right)\frac{1}{\epsilon}
    \right]
    +C_A^2
    \left[
      -\frac{1}{48 \epsilon ^2}
      +
      \left(
        \frac{17}{324}
        +\frac{5}{36} l_2
        +\frac{1}{72} L_{\mu }
      \right) \frac{1}{\epsilon }
    \right]
    \nonumber \\ &&
    + T n_l
    \left[
      C_F
      \left(
        \frac{1}{27 \epsilon ^2}-\frac{29}{324 \epsilon }
      \right)
      + C_A
      \left(
        \frac{1}{108 \epsilon ^2}-\frac{53}{1296 \epsilon }
      \right)
    \right]
  \Biggr\}
  \,,\nonumber\\
  \tilde{Z}_p &=&
  1
  +\left(\frac{\alpha_s^{(n_l)}(\mu)}{\pi}\right)^2 \frac{C_F \pi^2}{\epsilon}
  \left(
    \frac{1}{8} C_A
    +\frac{1}{4} C_F
  \right)
  + \left(\frac{\alpha_s^{(n_l)}(\mu)}{\pi}\right)^3 C_F \pi^2
  \Biggr\{
    C_F^2
    \left[
       \frac{5}{144 \epsilon ^2}
      + \left(
         \frac{31}{144}
        -\frac{1}{2} l_2
        +\frac{5}{48} L_\mu
      \right) \frac{1}{\epsilon }
    \right]
    \nonumber \\ &&
    + C_F C_A
    \left[
      -\frac{5}{96 \epsilon ^2}
      + \left(
         \frac{199}{432}
        +\frac{1}{4} l_2
        +\frac{29}{96} L_{\mu }
      \right)\frac{1}{\epsilon }
    \right]
    +C_A^2
    \left[
      -\frac{1}{16 \epsilon ^2}
      + \left(
         \frac{2}{27}
        +\frac{1}{4} l_2
        +\frac{1}{24} L_{\mu }
      \right)\frac{1}{\epsilon }
    \right]
    \nonumber \\ &&
    +T n_l
    \left[
      C_F
      \left(
        \frac{1}{18 \epsilon ^2}-\frac{11}{108 \epsilon }
      \right)
      + C_A
      \left(
        \frac{1}{36 \epsilon ^2}-\frac{37}{432 \epsilon }
      \right)
    \right]
    + C_F T n_h \frac{1}{60 \epsilon }
  \Biggr\}
  \,.
\end{eqnarray}
Note that our method only provides numerical results for the pole
parts. However, the precision is sufficiently high such that the analytic
results can be reconstructed using the PSLQ algorithm~\cite{PSLQ}.

The renormalization constants are related to the anomalous dimensions via
\begin{eqnarray}
    \gamma_x = \frac{\mathrm{d} \ln (\tilde{Z}_x)}{\mathrm{d} \ln (\mu)} ~,
\end{eqnarray}
which leads to
\begin{eqnarray}
  \gamma_x =
  -4 \left( \frac{\alpha_s^{(n_l)}}{\pi} \right) ^2 \tilde{Z}_x^{(2,-1)}
  -6 \left( \frac{\alpha_s^{(n_l)}}{\pi} \right) ^3 \tilde{Z}_x^{(3,-1)}
  + {\cal O}(\alpha_s^4)
  ~,
\end{eqnarray}
where $\tilde{Z}_x^{(a,b)}$ denotes the contribution to $\tilde{Z}$ at
order $\alpha_s^a \epsilon^b$.

For the perturbative expansion of $c_x$ we set the renormalization scale of
the strong coupling constant to $\mu^2=m^2$ and write
\begin{eqnarray}
  c_x &=& 1 + \frac{\alpha_s^{(n_l)}(m)}{\pi} c_x^{(1)}
  + \left(\frac{\alpha_s^{(n_l)}(m)}{\pi}\right)^2 c_x^{(2)}
%  \nonumber\\&&\mbox{}
  + \left(\frac{\alpha_s^{(n_l)}(m)}{\pi}\right)^3 c_x^{(3)}
  + {\cal O}(\alpha_s^4)
  \,.
  \label{eq::cxdef}
\end{eqnarray}
The three-loop coefficient is further decomposed
according to the color structures as
\begin{eqnarray}
%  \lefteqn{c_x^{(3)} =}
  c_x^{(3)} &=&
%  \nonumber\\&&\mbox{}
  C_F \big[ C_F^2 c^x_{FFF} + C_F C_A c^x_{FFA} + C_A^2 c^x_{FAA}
%  \nonumber\\&&\mbox{}
  + T n_l\left(
  C_F c^x_{FFL} + C_A c^x_{FAL} + T n_h c^x_{FHL} + T n_l c^x_{FLL}
  \right)
  \nonumber\\&&\mbox{}
  + T n_h\left(
  C_F c^x_{FFH} + C_A c^x_{FAH} + T n_h c^x_{FHH}
  \right) \big]
%'  \nonumber\\&&\mbox{}
  + \mbox{singlet terms}
  \label{eq::cv3ldef}
  \,.
\end{eqnarray}

In the following we present result for $c_x$ where for completeness also the
one- and two-loop results are shown. For the vector current our results read:
\begin{eqnarray}
  c_v^{(1)}&=&-2C_F\,,
  \nonumber\\
  c_v^{(2)}&=&\left(-\frac{151}{72}
    +\frac{89}{144}\pi^2
    -\frac{5}{6}\pi^2l_2-\frac{13}{4}\zeta(3)\right)C_AC_F
%  \nonumber\\&&\mbox{}
  +\left(\frac{23}{8}-\frac{79}{36}\pi^2
    +\pi^2l_2-\frac{1}{2}\zeta(3)\right)C_F^2
  \nonumber\\&&\mbox{}
  +\left(\frac{22}{9}-\frac{2}{9}\pi^2\right)C_FTn_h
  +\frac{11}{18}C_FTn_l
%  \nonumber\\&&\mbox{}
  -\frac{1}{2}\pi^2\left(\frac{1}{2}C_A
    + \frac{1}{3}C_F\right)C_F\Lmu
  \,,
  \nonumber\\
  c^v_{FFF} &=& 36.49486246
  + \left( -\frac{9}{16} + \frac{3}{2} l_2 \right) \pi^2 \Lmu -
  \frac{5}{32} \pi^2 \Lmu^2
  \,,  \nonumber\\
  c^v_{FFA} &=& -188.0778417
  + \left( -\frac{59}{108} - \frac{3}{4} l_2 \right) \pi^2 \Lmu -
  \frac{47}{576} \pi^2 \Lmu^2
  \,,  \nonumber\\
  c^v_{FAA} &=& -97.73497327
  + \left( -\frac{2}{9} - \frac{3}{4} l_2 \right) \pi^2 \Lmu +
  \frac{1}{6} \pi^2 \Lmu^2
  \,,  \nonumber\\
  c^v_{FFL} &=& 46.69169291 + \frac{25}{108} \pi^2 \Lmu -
  \frac{1}{18} \pi^2 \Lmu^2
  \,, \nonumber\\
  c^v_{FAL} &=& 39.62371855 + \frac{37}{144} \pi^2 \Lmu -
  \frac{1}{12} \pi^2 \Lmu^2
  \,, \nonumber\\
  c^v_{FHL} &=& -\frac{557}{162} + \frac{26}{81} \pi^2\,,
  \nonumber\\
  c^v_{FLL} &=& -\frac{163}{162} - \frac{4}{27} \pi^2\,,
  \nonumber\\
  c^v_{FFH} &=& -0.8435622912 - \frac{1}{20} \pi^2 \Lmu\,,
  \nonumber\\
  c^v_{FAH} &=& -0.1024741615\,,
  \nonumber\\
  c^v_{FHH} &=& -\frac{427}{162} + \frac{158}{2835}\pi^2 +
  \frac{16}{9}\zeta(3) \,,
  \label{eq::cv3}
\end{eqnarray}
The coefficient of the logaritmic contributions and the coefficients
$c^v_{FHL}$ and $c^v_{FLL}$ have been reconstructed using our
numerical expressions.  They agree with the results presented in
Ref.~\cite{Marquard:2014pea}. Our numerical precision is
not sufficient to obtain the analytic expressions for $c^v_{FHH}$
which we take from Ref.~\cite{Marquard:2014pea}.
For all coefficients presented in numerical form
we have a precision of at least ten digits, which is
a significant improvement.
For example, for the non-fermionic coefficients the results
in Ref.~\cite{Marquard:2014pea} read
$c^v_{FFF}=36.55(0.53)$,
$c^v_{FFA}=-188.10(0.83)$ and
$c^v_{FAA}=-97.81(0.38)$.

For the remaining three currents we have
\begin{eqnarray}
  c_a^{(1)}&=& - C_F \,,
  \nonumber\\
  c_a^{(2)}&=& \left( -\frac{9}{8} \zeta(3) + \frac{35}{144} \pi^2 -\frac{101}{72} - \frac{7}{12} \pi^2 l_2 \right) C_A C_F
  + \left( -\frac{27}{16} \zeta (3) - \frac{9}{8} \pi^2 + \frac{23}{24} + \frac{19}{24} \pi^2 l_2 \right) C_F^2
  \nonumber\\&&\mbox{}
  + \left( \frac{20}{9} - \frac{2}{9} \pi^2 \right) C_F T n_h
  + \frac{7}{18} C_F T n_l
  + \pi^2 \left( -\frac{1}{12} C_A -\frac{5}{24} C_F \right) C_F \Lmu
  \,,
  \nonumber\\
  c^a_{FFF} &=& -4.764274486 + \left( - \frac{155}{288} + l_2 \right) \pi^2 \Lmu
  \,,  \nonumber\\
  c^a_{FFA} &=& -83.88648515 + \left( - \frac{1289}{1728} - \frac{1}{6} l_2 \right) \pi^2 \Lmu + \frac{115}{1152} \pi^2 \Lmu^2
  \,,  \nonumber\\
  c^a_{FAA} &=& -63.00619439 + \left( - \frac{17}{108} - \frac{5}{12} l_2 \right) \pi^2 \Lmu + \frac{1}{18} \pi^2 \Lmu^2
  \,,  \nonumber\\
  c^a_{FFL} &=& 28.13543651 + \frac{83}{432} \pi^2 \Lmu - \frac{5}{72} \pi^2 \Lmu^2
  \,, \nonumber\\
  c^a_{FAL} &=& 23.17119085 + \frac{53}{432} \pi^2 \Lmu - \frac{1}{36} \pi^2 \Lmu^2
  \,, \nonumber\\
  c^a_{FHL} &=& - \frac{415}{162} + \frac{20}{81} \pi^2\,,
  \nonumber\\
  c^a_{FLL} &=& - \frac{65}{162} - \frac{2}{27} \pi^2\,,
  \nonumber\\
  c^a_{FFH} &=& 0.8971357511\,,
  \nonumber\\
  c^a_{FAH} &=& -0.2169123942\,,
  \nonumber\\
  c^a_{FHH} &=& -0.01136428050
                \,,
  \label{eq::ca3}
                \\
%\end{eqnarray}
%\begin{eqnarray}
  c_s^{(1)}&=& - \frac{1}{2} C_F \,,
  \nonumber\\
  c_s^{(2)}&=& \left( -\frac{5}{4} \zeta (3) + \frac{1}{48} \pi^2 + \frac{49}{144} - \frac{1}{2} \pi^2 l_2 \right) C_A C_F
  + \left( -\frac{11}{4} \zeta (3) - \frac{37}{48} \pi^2 + \frac{5}{16} + \frac{1}{2} \pi^2 l_2 \right) C_F^2
  \nonumber\\&&\mbox{}
  + \left( \frac{121}{36} - \frac{1}{3} \pi^2 \right) C_F T n_h
  - \frac{5}{36} C_F T n_l
  + \pi^2 \left( -\frac{1}{12} C_A -\frac{1}{3} C_F \right) C_F \Lmu
  \,,
  \nonumber\\
  c^s_{FFF} &=& -11.17444530 + \left( - \frac{53}{72} + l_2 \right) \pi^2 \Lmu
  \,,  \nonumber\\
  c^s_{FFA} &=& -83.13918787 + \left( - \frac{443}{432} - \frac{1}{6} l_2 \right) \pi^2 \Lmu + \frac{101}{576} \pi^2 \Lmu^2
  \,,  \nonumber\\
  c^s_{FAA} &=& -67.24288900 + \left( - \frac{17}{108} - \frac{5}{12} l_2 \right) \pi^2 \Lmu + \frac{1}{18} \pi^2 \Lmu^2
  \,,  \nonumber\\
  c^s_{FFL} &=& 30.10118322 + \frac{29}{108} \pi^2 \Lmu - \frac{1}{9} \pi^2 \Lmu^2
  \,, \nonumber\\
  c^s_{FAL} &=& 21.41321398 + \frac{53}{432} \pi^2 \Lmu - \frac{1}{36} \pi^2 \Lmu^2
  \,, \nonumber\\
  c^s_{FHL} &=& - \frac{157}{81} + \frac{5}{27} \pi^2
  \,,\nonumber\\
  c^s_{FLL} &=& \frac{73}{324} - \frac{1}{27} \pi^2
  \,,\nonumber\\
  c^s_{FFH} &=& 1.879249909
  \,,\nonumber\\
  c^s_{FAH} &=& -0.3740808359
  \,,\nonumber\\
  c^s_{FHH} &=& 0.007237324266
                \,,
  \label{eq::cs3}
                \\
%\end{eqnarray}
%\begin{eqnarray}
  c_p^{(1)}&=& -\frac{3}{2} C_F \,,
  \nonumber\\
  c_p^{(2)}&=& \left(- 3 \zeta (3) + \frac{17}{48} \pi^2 - \frac{17}{48} -\pi^2 l_2 \right) C_A C_F
  + \left( -\frac{9}{2} \zeta (3) - \frac{79}{48} \pi^2 + \frac{29}{16} + \pi^2 l_2 \right) C_F^2
  \nonumber\\&&\mbox{}
  + \left( \frac{43}{12} - \frac{1}{3} \pi^2 \right) C_F T n_h
  + \frac{1}{12} C_F T n_l
  + \pi^2 \left( -\frac{1}{4} C_A - \frac{1}{2} C_F \right) C_F \Lmu
  \,,
  \nonumber\\
  c^p_{FFF} &=& -16.65729478 + \left( \frac{5}{48} + \frac{3}{2} l_2 \right) \pi^2 \Lmu - \frac{5}{32} \pi^2 \Lmu^2
  \,,  \nonumber\\
  c^p_{FFA} &=& -181.0487647 + \left( - \frac{145}{144} - \frac{3}{4} l_2 \right) \pi^2 + \frac{1}{192} \pi^2 \Lmu^2
  \,,  \nonumber\\
  c^p_{FAA} &=& -104.3591595 + \left( - \frac{2}{9} - \frac{3}{4} l_2 \right) \pi^2 \Lmu + \frac{1}{6} \pi^2 \Lmu^2
  \,,  \nonumber\\
  c^p_{FFL} &=& 51.93841187 + \frac{11}{36} \pi^2 \Lmu - \frac{1}{6} \pi^2 \Lmu^2
  \,, \nonumber\\
  c^p_{FAL} &=& 39.92104383 + \frac{37}{144} \pi^2 \Lmu - \frac{1}{12} \pi^2 \Lmu^2
  \,, \nonumber\\
  c^p_{FHL} &=& - \frac{76}{27} + \frac{7}{27} \pi^2
  \,,\nonumber\\
  c^p_{FLL} &=& - \frac{41}{108} - \frac{1}{9} \pi^2
  \,,\nonumber\\
  c^p_{FFH} &=& 3.081762039 - \frac{1}{20} \pi^2 \Lmu
  \,,\nonumber\\
  c^p_{FAH} &=& -0.8953812450
  \,,\nonumber\\
  c^p_{FHH} &=& 0.06984121227
                \,.
  \label{eq::cp3}
\end{eqnarray}
For the axial-vector, scalar and pseudo-scalar current the terms
proportional to $n_l$ and $n_l^2$ can be found in Ref.~\cite{Piclum:2007an}.
There, the non-logarithmic terms of the coefficients $c^x_{FFL}$ and $c^x_{FAL}$
only have a precision of two significant digits whereas we have 
a precision of at least ten digits. Our analytic results for
$c^x_{FHL}$ and $c^x_{FLL}$ agree with~\cite{Piclum:2007an}.

After specifying the number of colours to
three we have for $\mu^2=m^2$ and $n_h=1$
\begin{eqnarray}
  c_v &\approx& 1 - \frac{\alpha_s^{(n_l)}}{\pi} \cdot 2.66667
  + \left(\frac{\alpha_s^{(n_l)}}{\pi}\right)^2\left[
    - 44.5510 + 0.407407 \, n_l \right]
  + \left(\frac{\alpha_s^{(n_l)}}{\pi}\right)^3\left[
    - 2090.33  + 120.661 \, n_l
%  \right.\nonumber\\&&\left.\mbox{}
    - 0.822779\, n_l^2
    \right]
  \nonumber\\&&\mbox{} + \mbox{singlet terms}
  \,,
\nonumber\\
  c_a &\approx& 1 - \frac{\alpha_s^{(n_l)}}{\pi} \cdot 1.33333
  + \left(\frac{\alpha_s^{(n_l)}}{\pi}\right)^2\left[
    - 29.3816 + 0.259259 \, n_l \right]
  + \left(\frac{\alpha_s^{(n_l)}}{\pi}\right)^3\left[
    - 1214.40  + 71.3101 \, n_l
    - 0.377439\, n_l^2
    \right]
  \nonumber\\&&\mbox{} + \mbox{singlet terms}
  \,, \nonumber\\
  c_s &\approx& 1 - \frac{\alpha_s^{(n_l)}}{\pi} \cdot 0.666667
  + \left(\frac{\alpha_s^{(n_l)}}{\pi}\right)^2\left[
    - 30.2266 - 0.0925926 \, n_l \right]
  + \left(\frac{\alpha_s^{(n_l)}}{\pi}\right)^3\left[
    - 1275.89 + 69.5462 \, n_l
    - 0.0467441 \, n_l^2
    \right]
  \nonumber\\&&\mbox{} + \mbox{singlet terms}
  \,, \nonumber\\
  c_p &\approx& 1 - \frac{\alpha_s^{(n_l)}}{\pi} \cdot 2
  + \left(\frac{\alpha_s^{(n_l)}}{\pi}\right)^2\left[
    - 52.1381 + 0.0555556 \, n_l \right]
  + \left(\frac{\alpha_s^{(n_l)}}{\pi}\right)^3\left[
    - 2256.42 + 125.924 \, n_l
    - 0.492084 \, n_l^2
    \right]
  \nonumber\\&&\mbox{} + \mbox{singlet terms} \,.
  \label{eq::c3num}
\end{eqnarray}
For all four currents the quantum corrections are quite sizable. For
applications in the top quark sector, i.e.\ for $n_l=5$, the two- and three-loop
corrections have the same order of magnitude as the one-loop term. For $n_l=3$
and $n_l=4$ the higher order corrections are even larger. Since the matching
coefficients on their own are no physical quantities this is no principle
problem.  However, it shows the importance of the three-loop corrections to $c_x$, in
particular for $c_v$ which has important applications in the
bottom~\cite{Beneke:2014qea,Egner:2021lxd} and top
sector~\cite{Beneke:2015kwa}.

%- }}}
%- {{{ Concl.:

\section{\label{sec::concl}Conclusions}

In this work we have computed the three-loop corrections to the QCD-NRQCD
matching coefficients for external vector, axial-vector, scalar and
pseudo-scalar currents. We consider the corresponding quark form factors and
compute the pure-hard part of each master integral using the method of
Ref.~\cite{Fael:2021kyg} supplemented with the information from expansions by
regions~\cite{Beneke:1997zp}. We obtain precise numerical results for the
three-loop coefficients. For the vector current we provide the first
independent cross check for $c_v$ which has a significant numerical impact to
the N$^3$LO predictions for top-quark-pair production in electron-positron
annihilation close to threshold and the leptonic decay width of the 
$\Upsilon(1S)$ meson.
Our new result is several orders of magnitude
more precise. The three-loop results for $c_a$, $c_s$ and $c_p$ are new.

%- }}}

\bigskip

{\bf Acknowledgements.}
This research was supported by the Deutsche Forschungsgemeinschaft (DFG,
German Research Foundation) under grant 396021762 --- TRR 257 ``Particle
Physics Phenomenology after the Higgs Discovery''.  The Feynman diagrams were
drawn with the help of Axodraw~\cite{Vermaseren:1994je} and
JaxoDraw~\cite{Binosi:2003yf}.

%\begin{appendix}
%
%\section*{Appendix ???}
%
%\end{appendix}

\end{document}